\documentclass[fleqn,usenatbib]{mnras}

\usepackage{newtxtext,newtxmath}

\usepackage[T1]{fontenc}

\DeclareRobustCommand{\VAN}[3]{#2}
\let\VANthebibliography\thebibliography
\def\thebibliography{\DeclareRobustCommand{\VAN}[3]{##3}\VANthebibliography}

\usepackage{graphicx}	
\usepackage{amsmath}	
\usepackage{caption}
\usepackage{anyfontsize}
\usepackage{pdflscape}
\usepackage{afterpage}
\usepackage{subcaption}

\DeclareUnicodeCharacter{2212}{-}

\makeatletter
 \def\@textbottom{\vskip \z@ \@plus 9pt}
 \let\@texttop\relax
\makeatother

\title[Sub-Minute Optical Transients with DWF]{A Machine Learning empowered search for Sub-Minute Optical Transient Events with the \textit{Deeper, Wider, Faster} programme}

\author[S. Goode et al.]{Simon R. Goode,$^{1,2,3}$
Sara A. Webb,$^{1,2}$\thanks{E-mail: swebb@swin.edu.au}
Jeff Cooke,$^{1,2}$ 
Jielai Zhang,$^{1,2}$
James Freeburn,$^{1,2}$
\newauthor
Amy Lien,$^{4}$
Mohsen Shamohammadi,$^{1,2}$
Alexandra Rosenthal,$^{1}$
Laura N. Driessen,$^{2,7}$
\newauthor
Christopher Fluke,$^{1}$
Ashish Mahabal,$^{5,6}$
Anais Möller,$^{1,2}$
Dougal Dobie,$^{2,7}$
\newauthor
Adam Batten,$^{1}$
Natasha Van Bemmel$^{1,2}$
\\
$^{1}$Centre for Astrophysics and Supercomputing, Swinburne University of Technology, Hawthorn, Melbourne 3122, Australia\\
$^{2}$OzGrav: The ARC Centre of Excellence for Gravitational Wave discovery, Swinburne University of Technology, Hawthorn, Melbourne 3122, Australia\\
$^{3}$School of Physics and Astronomy, Monash University, VIC 3800, Australia\\
$^{4}$University of Tampa, Department of Physics and Astronomy, 401 W. Kennedy Blvd, Tampa, FL 33606, USA\\
$^{5}$Division of Physics, Mathematics and Astronomy, California Institute of Technology, Pasadena, CA 91125, USA\\
$^{6}$Center for Data Driven Discovery, California Institute of Technology, Pasadena, CA 91125, USA\\
$^{7}$Sydney Institute for Astronomy, School of Physics, University of Sydney, Sydney, NSW 2006, Australia\\
}

\date{Accepted XXX. Received YYY; in original form ZZZ}

\pubyear{2025}

\begin{document}
\label{firstpage}
\pagerange{\pageref{firstpage}--\pageref{lastpage}}
\maketitle

\begin{abstract}
Optical transient surveys continue to generate increasingly large datasets, prompting the introduction of machine-learning algorithms to search for quality transient candidates efficiently. Existing machine-learning infrastructure can be leveraged in novel ways to search these datasets for new classes of transients. We present a machine-learning accelerated search pipeline for the Deeper, Wider, Faster (DWF) programme designed to identify high-quality astrophysical transient candidates that contain a single detection. Given the rapid observing cadence of the DWF programme, these single-detection transient candidates have durations on sub-minute timescales. This work marks the first time optical transients have been systematically explored on these timescales, to a depth of m$\sim$23. We report the discovery of two high-quality sub-minute transient candidates from a pilot study of 671,761 light curves and investigate their potential origins with multiwavelength data. We discuss, in detail, possible non-astrophysical false positives, confidently reject electronic artefacts and asteroids, ruling out glints from satellites below 800 km and strongly disfavouring those at higher altitudes. We calculate a rate on the sky of $4.72^{+6.39}_{-3.28}\times10^5$ per day for these sub-minute transient candidates.
\end{abstract}




\section{Introduction}
The recent discoveries of Fast Radio Bursts \citep[FRBs;][]{Lorimer2007, Petroff2022} and kilonovae \citep[KNe;][]{Metzger2019, Abbott2017} have placed growing value on transient surveys and counterpart searches with wide fields of view, deep limiting magnitudes and fast-cadenced imaging.
Detection pipelines for fast-transient events benefit from data-driven survey strategies that probe large cosmological volumes (to find more of these rare events) at a high temporal resolution (to better understand their evolution and physical properties).
In addition, surveys that can quickly trigger follow-up or simultaneously observe with multiwavelength facilities have the added opportunity to detect multiwavelength counterparts, leading to a better understanding of the physical properties of their progenitor systems. 

The Deeper, Wider, Faster programme (DWF) is one such transient survey, which utilises deep, wide-field fast-cadenced imaging and both simultaneous and rapid-response follow-up triggers of multiwavelength facilities \citep[Cooke et al. in prep.;][]{Andreoni2018, Andreoni2020, Webb2020, Webb2021, Freeburn2024, freeburn2025}.
This enables the DWF programme to target fast-transient phenomena, typically events with millisecond-to-hours duration.
The DWF programme takes continuous 20-second exposures, which, including CCD readout time, result in minute-cadence images.

Searching through transient candidates that only have one detection is fraught with challenges.
Most of such candidates are likely to be artefacts caused by cosmic rays, electronic readout cross-talk or poor subtraction residuals.
Thankfully, machine-learning (ML) algorithms have recently made strides in transient astronomy, and many are designed specifically to address this problem.
`Real/Bogus' (or RB) classification algorithms \citep[e.g.][]{Bailey2007, Bloom2012, Wright2015, Masci2017, Duev2019, Mong2020, Hosenie2021, Killestein2021, Chang2021, Makhlouf2022, Takahashi2022, Mong2023, Acero-Cuellar2023, Weston2024, Pan2024, Liu2025, Semenikhin2025, Gu2025, Shi2025} are designed to inspect an image of a transient candidate (typically the transient \textit{subtraction} or \textit{difference} image) and classify whether or not the detection is of astrophysical origin.

In this paper, we use the DWF programme to probe a relatively unexplored regime of optical transients, those with characteristic timescales of one minute or less.
To search for these transients, we develop a method to identify high-likelihood transient candidates that contain a single detection.
Since we search for single-detection candidates, we inevitably sacrifice any temporal information about the source.
We can, however, leverage the superior depth of the Dark Energy Camera \citep[DECam;][]{Flaugher2015} located at the Cerro Tololo Inter-American Observatory $g$-band ($\sim$23-mag with 20s exposure time) to probe as large a cosmological volume as possible.
We utilise the DWF-designed real/bogus classification algorithm from the Removal of Bogus Transients (\textsc{robot}) pipeline \citep{Goode2022} to efficiently filter and identify sub-minute transients.  

We summarise the DWF programme and outline the data collection process in \S\ref{sec:DWF}. In \S\ref{sec:SMOTEs}, we provide a comprehensive explanation of the sub-minute optical transient discovery pipeline, including selection criteria (\S\ref{subsec:selection_criteria}), image processing (\S\ref{subsec:candidate_processing}), calculation and use of parameters produced by \textsc{source extractor} and the \textsc{robot} pipeline (\S\ref{subsec:SE_robot}) and candidate filtering (\S\ref{subsec:filtering}). In \S\ref{sec:results_and_analysis} , we detail the findings of the search, including a search for multiwavelength counterparts (\S\ref{subsec:multiwavelength}), discussion of possible progenitors, estimates on the rates of sub-minute transient events, and their impact on the literature (\S\ref{subsec:discussion}). Finally, we provide concluding statements, advice and future work in \S\ref{sec:conclusions}.

\section{Data}
\label{sec:DWF}
The DWF programme is an all-wavelength (radio through $\gamma$-ray) and multimessenger survey designed to search for fast transients (milliseconds-to-days duration) in real-time.
DWF coordinates observations with multiwavelength and multimessenger facilities simultaneously and as rapid response follow-up.
For a given run, DWF coordinates $\sim$10 wide-field facilities across different wavelength regimes, processes the data near-real-time, and coordinates radio through $\gamma$-ray facilities for rapid-response and late-time follow-up observations.
Observing runs typically occur twice yearly and consist of 6 consecutive half-nights targeting 2 to 3 fields each. Here, we focus on the wide-field optical component of the programme, usually led with DECam or Hyper SuprimeCam \citep[HSC;][]{Aihara2017} in the optical \textit{g}-band to probe the sub-minute timescale.

To maximise the scientific value of fast transient discoveries, DWF utilises a 20-second exposure time when using DECam.
These 20-second exposures reach limiting magnitudes of $m(g)\sim23$ and enable a cadence of 1 minute (including CCD readout and clear time).
This strategy, usually in tandem with a stare observing pattern, provides excellent temporal resolution for minutes-to-hours duration transients whilst probing a large area of sky.
Moreover, this strategy affords the luxury of later co-addition to search for fainter objects at the cost of temporal resolution.
The short exposures and fast cadence also allow for easy identification of moving objects (e.g. artificial satellites, asteroids) in consecutive images.

The data used in this work is from the January 2015 pilot run of DWF, in which some observing strategies differ from the standard that has since emerged.
This pilot run utilised a dithering strategy to fill in CCD gaps.
This approach affords greater sky coverage at the cost of consistent temporal resolution. 
Raw data were processed by the NOAO Community Pipeline \citep{Valdes2014}.
The pipeline includes bias correction, crosstalk correction, bad pixel masking, flat field calibration, astrometric calibration and cosmic ray masking.
We note that the cosmic ray masking is not comprehensive, and some cosmic rays persist in the data at a reduced rate.
Individual CCDs were processed and sources were identified and extracted using \textsc{Source Extractor} \citep{Bertin1996}.
Zeropoint and magnitude offset corrections were obtained from the SkyMapper DR2 catalogue \citep{Onken2019}.
Refer to \citet{Webb2020} for more information on data processing and lightcurve construction.

We limit the contamination rate caused by artificial satellites and debris as the data was collected before the launch of the SpaceX Starlink mega constellation (first 60 satellites launched in May 2019) \citep{Tyson2020, McDowell2020}.
We note that the available multiwavelength coverage during this run was considerably less than current standards.
Future work in this area will process all the DWF archival data, including runs where radio, UV, X-ray, $\gamma$-ray and high-energy particle data were taken simultaneously.
The data used in this study were generated for the previous works \cite{Webb2020, Webb2021}, which searched for optical transients as short as three minutes.

Previous works have searched for extragalactic fast transients in the archival DWF data \citep{Freeburn2024, freeburn2025, Andreoni2020} and achieved rates and upper limits.
These works considered events with 3 or more detections to help confirm the reality of their sources.
In this work, we use DECam data obtained for the DWF programme simultaneously with Murriyang, CSIRO's Parkes radio telescope, the Molonglo Observatory Synthesis Telescope (MOST), and with follow-up from the Neil Gehrels Swift Observatory \citep{Gehrels2004}.

The Murriyang (Parkes) radio telescope is a 64 m single dish antenna and, at the time of observation, had the 13-beam 21 cm multibeam receiver mounted \citep{Staveley1996}.
The Murriyang data were recorded as part of the P858 observing program using the multibeam receiver with the Berkeley-Parkes-Swinburne Recorder (BPSR) processor backend \citep{McMahon2008, Keith2010} with paralactic angle tracking enabled (to remain on field throughout the observation).
These data spans a bandwidth of approximately 400\,MHz centred at 1382\,MHz.

MOST was a Mills-Cross interferometer comprising two fully steerable east-west 778m long arms, with a total collecting area of 18,000m$^{2}$.
During this observation run, MOST was undergoing its transformation into the UTMOST upgraded facility to search for FRBs \citep{Caleb2017, Bailes2017, Jankowski2019}.

The fields targeted during the selected observational run can be seen in Table \ref{tab:field_info}.
Overall, the data used in this work amounts to 671,763 lightcurves across 2 hours and 7 minutes of consecutive, 20s exposures on-sky.

\begin{table}
    \centering
    \caption{Summary of fields, light curves, total exposure times, and field centres for the data included in this work.}
    \label{tab:field_info}
    \begin{tabular}{lcccr}
        \hline
        Field & Light Curves & Total Exposure Time & $\alpha$ & $\delta$ \\ \hline
        CDF-S & 144,499 & 54m 20s & 3h32m28s & −27$^\circ$48'30" \\
        4hr & 527,264 & 1h 12m 40s & 4h10m00s & −55$^\circ$00'00" \\
        Total & 671,763 & 2h 7m & - & - \\ \hline
    \end{tabular}
\end{table}

\section{Analysis}
\label{sec:SMOTEs}

\subsection{Light Curve Selection Criteria}
\label{subsec:selection_criteria}

\begin{figure}
    \centering
    \includegraphics[width=\columnwidth]{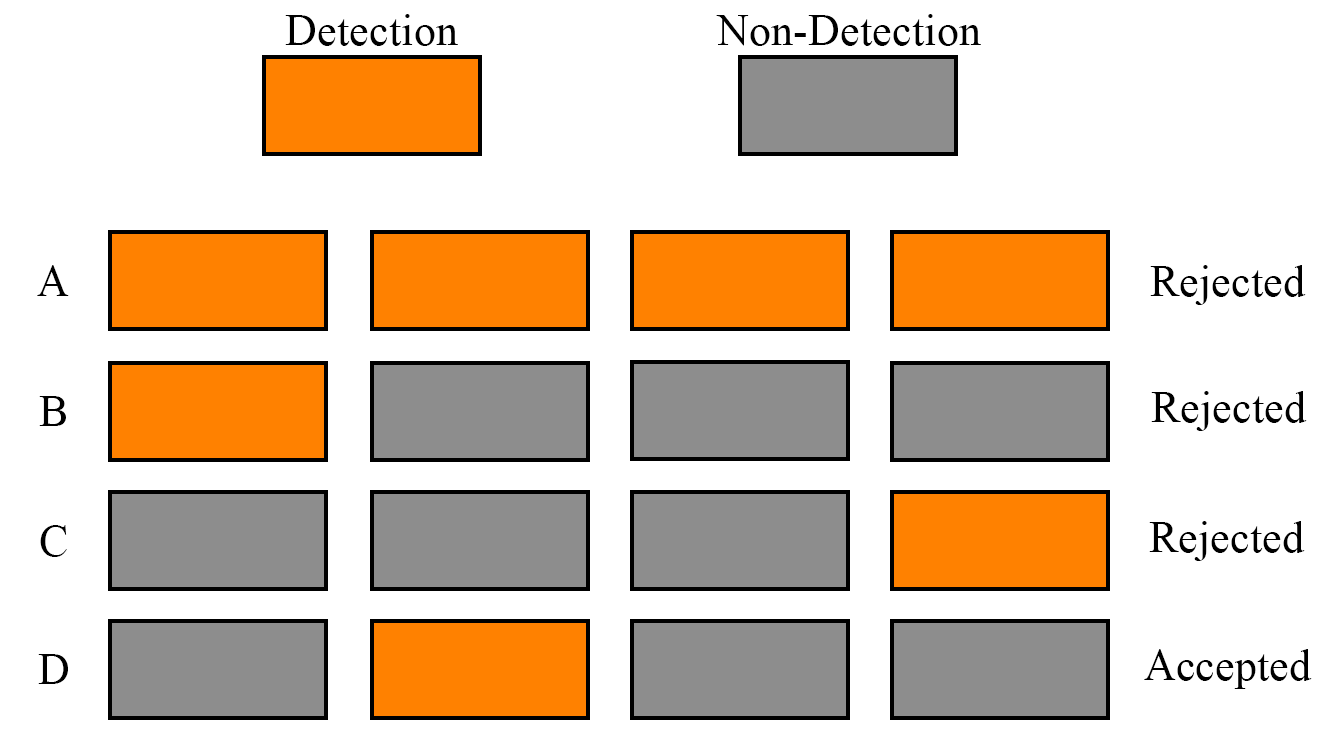}
    \caption{Example of selection criteria for the sub-minute optical transient candidates. Given a set of consecutive images (four images in this example; our data consists of $\sim$100 images per night), we define the following selection criteria: single detection, present anywhere except the first and last image.}
    \label{fig:smoteselection}
\end{figure}

The selection criteria for DECam light curves in transient pipelines determine which light curves are comprehensively analysed.
In our work, these criteria are deliberately conservative to minimise the risk of excluding potential true positives from the data.
We are explicitly searching the light curves for sources not seen in the template image, that may or may not have a host galaxy.
Therefore, the selected light curves must have a single detection identified by \textsc{source extractor}, given a \textsc{detect\_thresh} parameter set at 1.1, where the other data points in the light curve are non-detections.
If the single detection is the first or last of the light curve, we cannot verify that the detections immediately before/after the single detection are above the threshold.
Figure \ref{fig:smoteselection} helps to visualise the selection criteria on the transient duration. In these cases, the light curve fails the criteria and is skipped. 
These criteria aim to find the following: 
\begin{enumerate}
    \item Sub-minute bursts that have faint or non-visible (i.e. below the limiting magnitude) quiescent counterparts, such as transients that originate from the outskirts of a visible extragalactic host (i.e. transient source is not coupled with host galaxy core)
    \item Transients that originate near the core of a faint extragalactic host whose brightness is below the limiting magnitude
    \item Transients with no apparent host galaxy (e.g. a Galactic transient such as a flare from a faint star)
\end{enumerate} 
Using these criteria, we reduce the number of light curves across both fields from 671,761 to 385,775 sub-minute optical transient candidates (hereafter referred to as candidates). 

\subsection{Candidate Image Processing}
\label{subsec:candidate_processing}
Light curves that meet the selection criteria are moved into the image-processing phase of the pipeline to assess source profile and characteristics.
This phase is crucial as many known artefacts, such as cosmic rays and electronic artefacts, can appear as false positives when relying solely on a single detection in the light curves. 

The processing begins by identifying where and when the candidate occurred, namely on which DECam CCD and during which exposure.
The exposure is identified from the light curve, and the correct CCD is identified by cross-matching the source coordinates within the CCD bounds.
With the candidate detections identified, a suitable template image of the field must be selected for further subtraction in the pipeline.
To ensure that all templates adequately reflect the observation conditions of the candidates (e.g. seeing, air mass, weather effects), templates are stacked from exposures taken on the same night as the candidate, as near to the time of the candidate as possible, including a buffer period of 3 minutes before and after the candidate detection.
The buffer period ensures that any small changes in magnitude immediately before/after the candidate do not contaminate the template image. Finally, the template images are stacked and aligned with the candidate image using the \textsc{swarp} software package \citep{Bertin2002}.

Once the template has been stacked and aligned with the candidate image, the template is subtracted from the candidate image to produce a residual subtraction (or `difference') image using the \textsc{hotpants} software package \citep{Becker2015}. This subtraction image effectively highlights any objects in the candidate image that brighten or fade with respect to the template image by showing positive or negative residual flux. The production of a subtraction image is necessary to aid manual candidate inspection efforts, as well as to flag bogus candidates. 

\subsection{Sub-minute Transient Candidate Selection \& Artefact Filtering}
\label{subsec:filtering}
\label{subsec:SE_robot}
\begin{figure}
\centering
\includegraphics[width=\columnwidth]{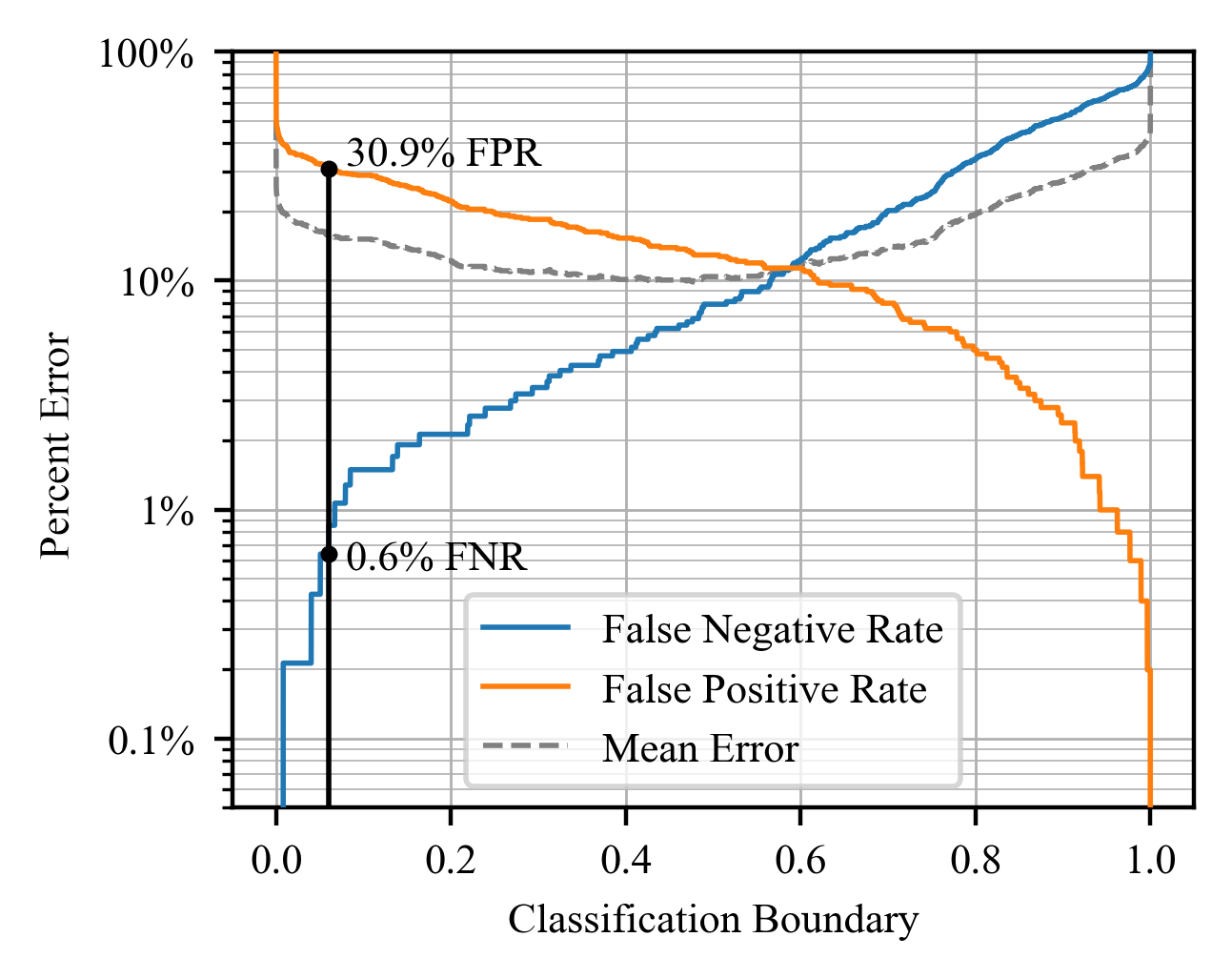}
\caption{Misclassification performance of the \textsc{robot} CNN classifier as a function of the decision threshold. At the decision boundary of 0.06 (vertical black line), the algorithm performs with a 0.6\% False Negative Rate (FNR) and 30.9\% False Positive Rate (FPR). Since we are considering data with a large bias towards very low scores ($<$0.02) and are looking for a rare class of transient, our objective is to minimise the FNR as much as is feasible without diluting the results with too many contaminants.}
\label{fig:robot_boundary}
\end{figure}

Here, we strive to find PSF-like point source transients, the signal we expect from a genuine astrophysical transient.
We utilise the Convolutional Neural Network (CNN) built for the \textsc{robot} pipeline \citep{Goode2022}, which was trained to detect PSF-like point sources in a data-driven manner.
During training, this algorithm learned to extract informative features from the template, science and subtraction images simultaneously.
The \textsc{robot} CNN has been demonstrated to confidently detect transient events in $g$-band images taken by DECam.
The extracted features assess different aspects of the shape of a source and can be used to judge their PSF-like qualities.
We acknowledge that using this preferential search may miss very fast bursts that do not produce PSF-like profiles, whose durations are sufficiently shorter than the timescale needed for atmospheric distortions to produce PSF blurring effects.
Such short-duration bursts are expected to be seen as speckled images, and detections of these patterns are the subject of future work. 

As with all binary classifiers, a decision boundary must be selected for the task of sorting real candidates from bogus artefacts.
Figure \ref{fig:robot_boundary} shows the false negative rate, false positive rate and the mean misclassification error of the \textsc{robot} CNN, as a function of the decision boundary.
For this work, we elect a low decision boundary of 0.06 to remain conservative when filtering candidates.
With a decision boundary at 0.06, Figure \ref{fig:robot_boundary} shows the real/bogus algorithm to have a false negative rate (FNR) of 0.6\%.
With an FNR of only 0.6\%, our model achieves a completeness of 99.4\%, retaining 99.4\% of all genuine transients in the filtered candidate dataset. 
Although such a low decision boundary will introduce false-positive contaminants (FPR > 30\%), it is preferable to minimise the FNR, as the cost of losing a genuine transient is far higher than the cost of manually vetting a false positive.
Applying this decision boundary to the full set of 385,775 candidates filtered it down to 5,477.
This ML-enabled filtration step removed $>$98\% of the candidates, while retaining a hypothetical true positive rate of $>$99\%. 

We explored the use of unsupervised clustering techniques for the discovery of candidates, which ultimately proved that the most impactful feature was the real/bogus score of each candidate. This work can be seen in Appendix \ref{app:SE_robot}.

\section{Results}
\label{sec:results_and_analysis}

\begin{figure*}
    \centering
    \includegraphics[width=\linewidth]{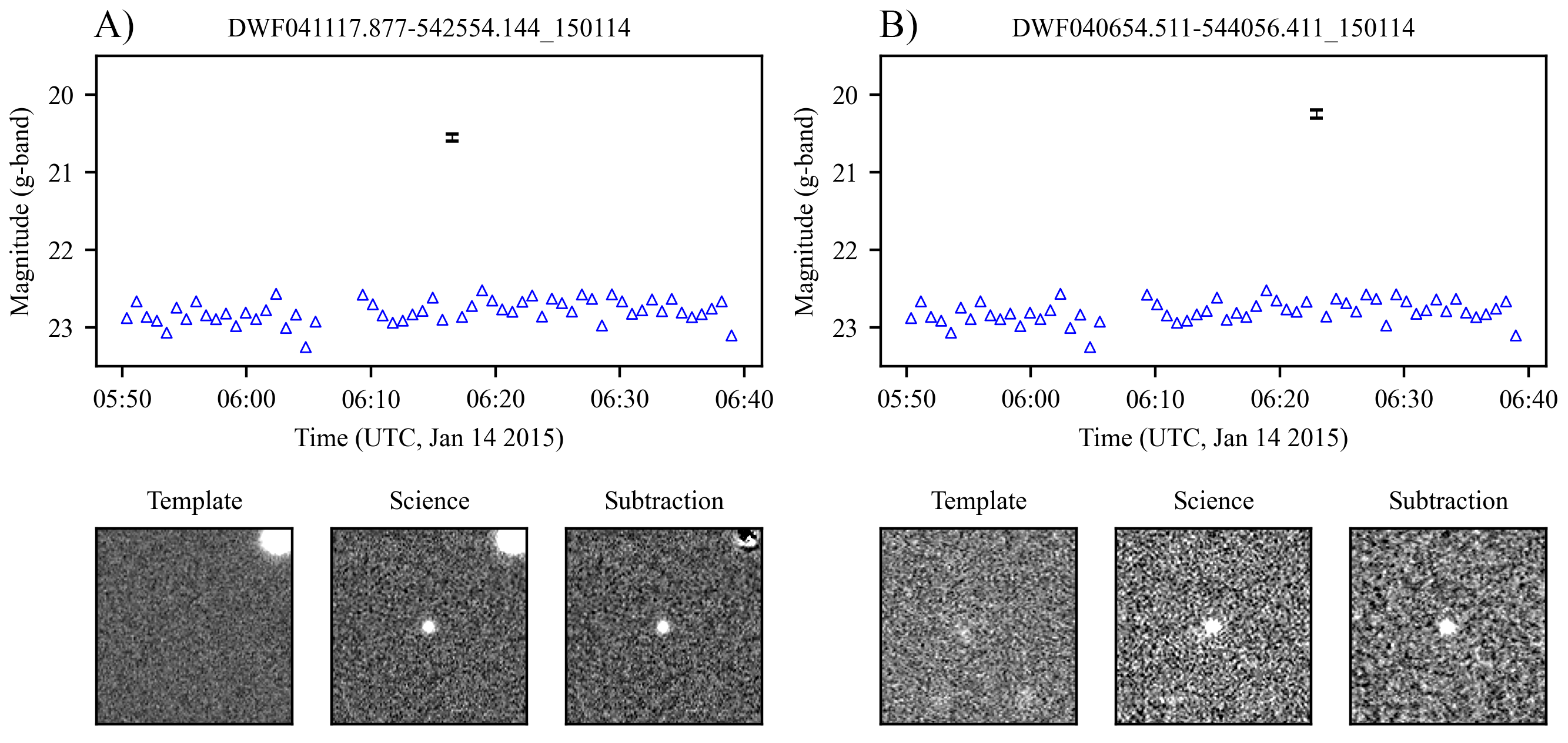}
    \caption{Light curves and detection images for the two high-confidence candidates. A) is candidate DWF041117.877-542554.144, and B) is candidate DWF040654.511-544056.411. The light curves, upper row, display the DECam $g$-band non-detection upper limits as blue triangles, and the single detection apparent magnitude as the black point. The detection images, bottom row (left to right), display the template image for that region of sky taken on the observational night, three minutes prior to the science detection, the science image of the single detection, and the subtraction of the science image from the template, leaving the residual flux from the source.}
    \label{fig:2cands}
\end{figure*}

\begin{figure}
    \centering
    \includegraphics[width=\columnwidth]{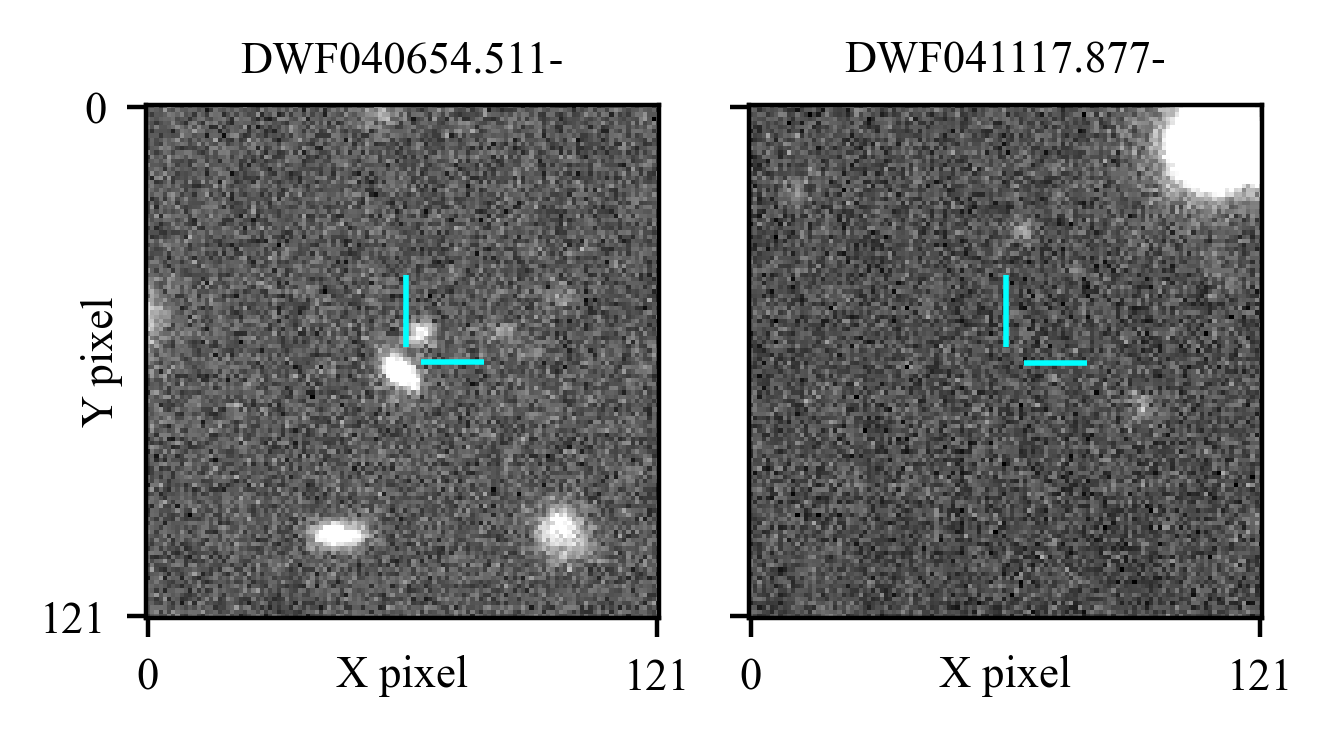}
    \caption{Deep field $g$-band imaging of the sky region surrounding the sub-minute candidates, DWF040654.511- (left) and DWF041117.877 (right). These deep field images are a stack of one night of data and reach a limiting magnitude of $m(g)\sim25$.}
    \label{fig:deepstack}
\end{figure}

After the filtration process, the remaining 5,477 candidates were manually inspected by multiple members of the DWF team to search for the most realistic and promising transients.
This search yielded two high-confidence candidates, DWF040654.511-544056.411 and DWF041117.877-542554.144, which we report in this work.
The \textit{g}-band light curves, templates, subtractions and science images of the two candidates are presented in Figure \ref{fig:2cands}.
In addition, the information for the two candidates, including \textsc{source extractor} measurements, \textsc{robot} score and calculations of absolute magnitude (detailed later in this section), is provided in Table \ref{tab:candetails}.
The data inspectors noted that the remaining 5,475 candidates included contaminants (i.e. appropriately low-scoring artefacts), misclassifications (i.e. artefacts with mid-to-high scores) and ambiguous or otherwise uninteresting objects (e.g. misaligned subtractions, satellite streaks).

Of the two promising candidates, both were found in the 4hr field. DWF040654.511-544056.411 appears to have a host galaxy in deeper field imaging (shown in Figure \ref{fig:deepstack}), while DWF041117.877-542554.144 appears hostless.

\begin{figure}
\includegraphics[width=\columnwidth]{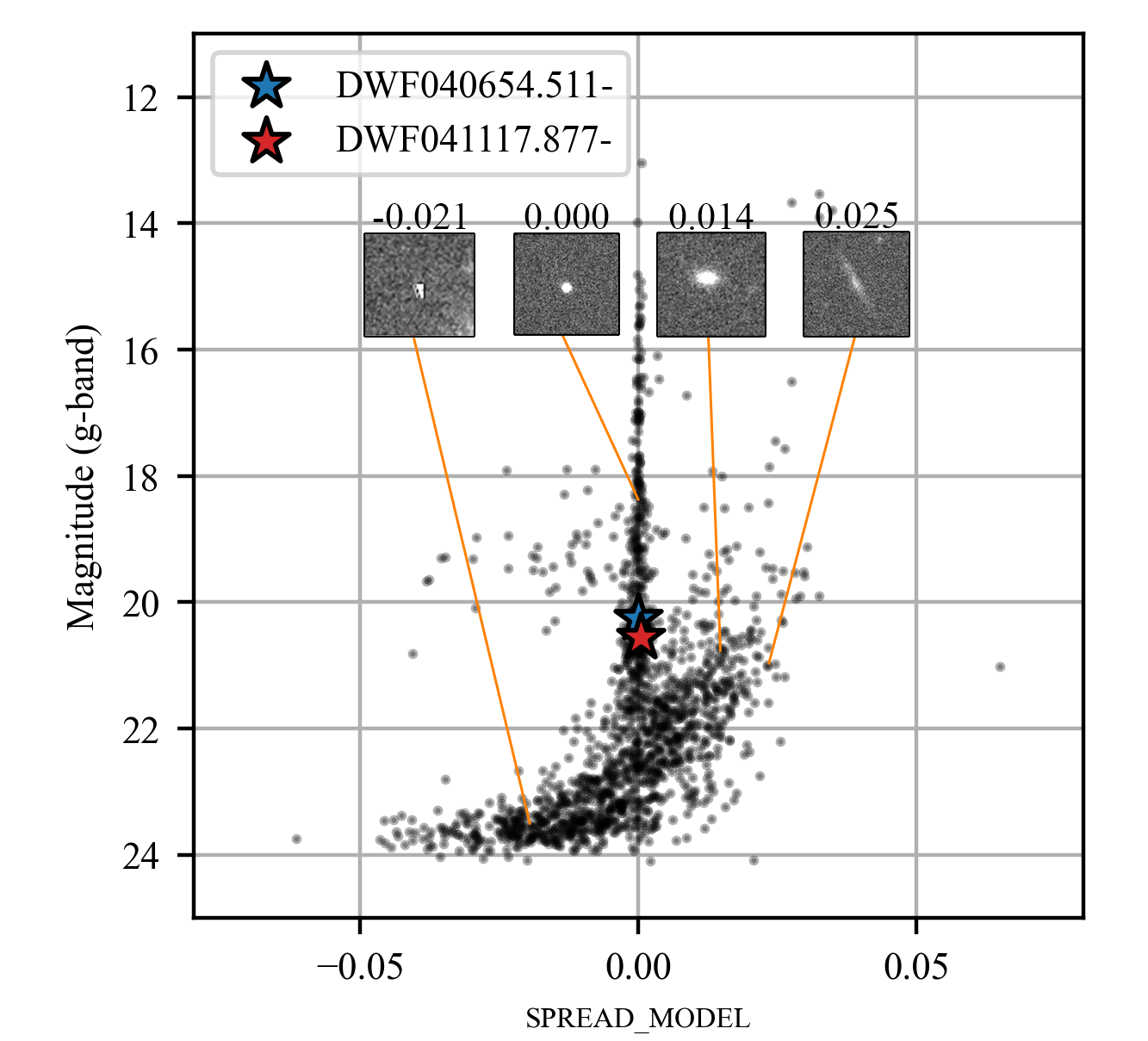}
\caption{Point Spread Function (PSF) analysis of the two promising sub-minute optical transients candidates. This diagnostic plot shows the \textsc{source extractor} \textsc{spread\_model} of sources found on promising candidate CCDs, with the 2 promising candidates highlighted as stars. \textsc{spread\_model} is defined by Equation \ref{eq:sm} and is used as a star/galaxy classifier. A stellar locus (PSF-like sources) as a function of magnitude is featured at \textsc{spread\_model} values close to 0. Sources with \textsc{spread\_model} values greater than 0 outside of the stellar peak are classified as extended sources such as galaxies or nebulae, and those below 0 are classified as electronic artefacts or cosmic rays.}
\label{fig:spreadmodel}
\end{figure}

\begin{figure}
\includegraphics[width=\columnwidth]{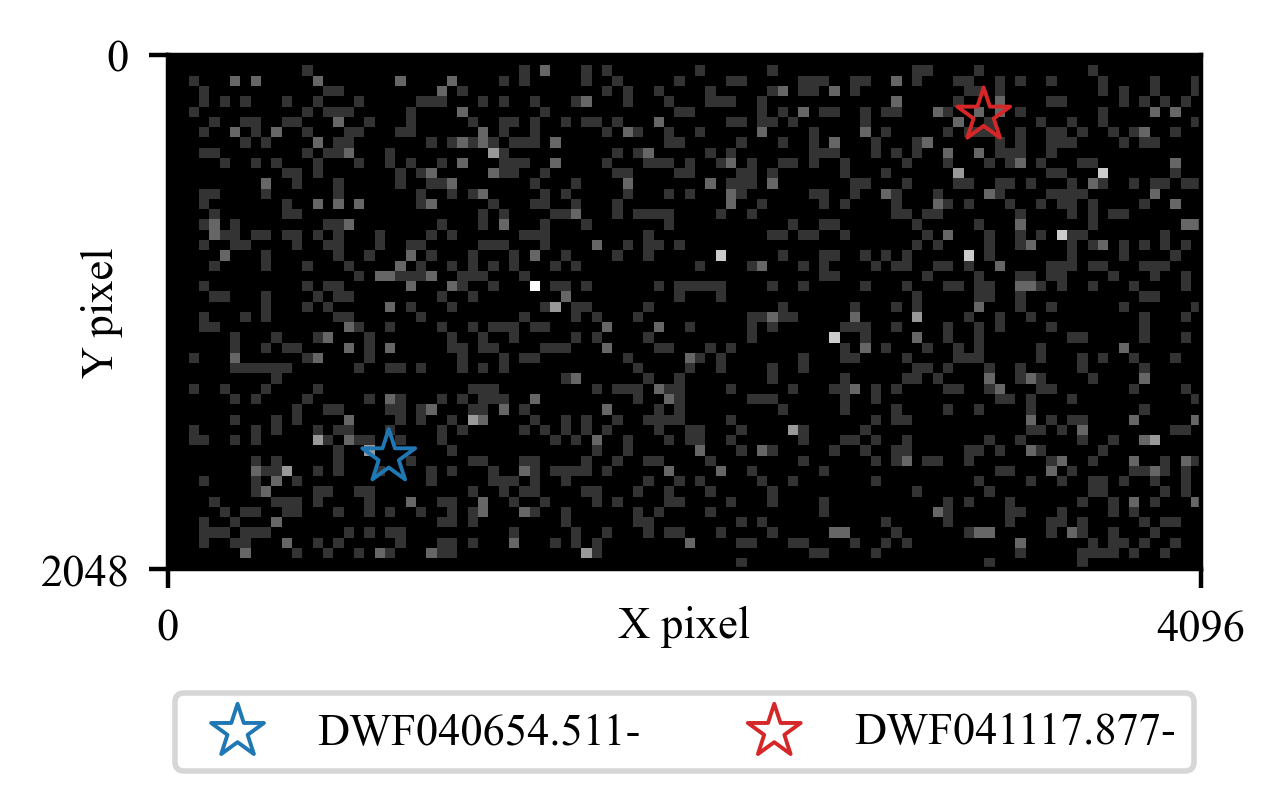}
\caption{CCD pixel location diagnostic plot of sources found on their respective candidate CCDs, with the two high-confidence candidates highlighted as stars. Pixel location is a record of the x- and y-values at which the centre of a source is found. The locations of the two promising candidates appear random on their CCDs and do not reside close to CCD edges or straddle amplifier regions.}
\label{fig:xypos}
\end{figure}

For both candidates, we compile the information from all other sources on the CCD during the sub-minute optical transient timeframe.
Figure \ref{fig:spreadmodel} depicts the shapes of the two promising candidates using \textsc{spread\_model} compared to other sources on their respective CCDs and exposures.
\textsc{spread\_model} is an optional metric included within \textsc{source extractor}, designed to be used as a star/galaxy classifier \citep{Mohr2012}.
Specifically, \textsc{spread\_model} indicates whether a source in an image is best represented by a point source model ($\tilde\phi$) or a galaxy model ($\tilde G$).
\textsc{spread\_model} is defined as

\begin{equation}
\textsc{spread\_model} = \frac{\tilde G^TWp}{\tilde\phi^TWp} - \frac{\tilde G^TW\tilde\phi}{\tilde\phi^TW\tilde\phi}
\label{eq:sm}
\end{equation}
where $p$ is the image vector centred on the source, $W$ is a weight matrix (constant along the diagonal except for bad pixels, where the weight is 0), and a superscript $T$ indicates the transpose of the preceding vector.

By design, \textsc{spread\_model} values close to 0 are well described by the PSF model, positive \textsc{spread\_model} values are extended sources more appropriately described by the galaxy model, and negative values are likely cosmic rays or electronic artefacts with a FWHM smaller than the PSF model.
Figure \ref{fig:spreadmodel} shows a clear stellar locus at \textsc{spread\_model} values close to 0, where the two candidates lie.
Importantly, this figure demonstrates that neither of the sub-minute transient candidates has a morphology similar to cosmic rays or electronic artefacts.

It is common for electronic artefacts to appear along the edges and amplifiers of CCDs and affect images and lightcurves (see \citet{Webb2020, Goode2022} for examples).
Figure \ref{fig:xypos} compiles the CCD pixel positions of the two promising candidates compared to other sources in their respective exposures on the footprint of a 2K$\times$4K DECam CCD.
The two candidates are positioned far from the CCD edge and the horizontal division between the two amplifiers (along pixel 1024) of the DECam CCDs in the middle.
As the two candidates are sufficiently distant from those areas, they are unlikely to be such artefacts. 

\begin{figure}
\includegraphics[width=\columnwidth]{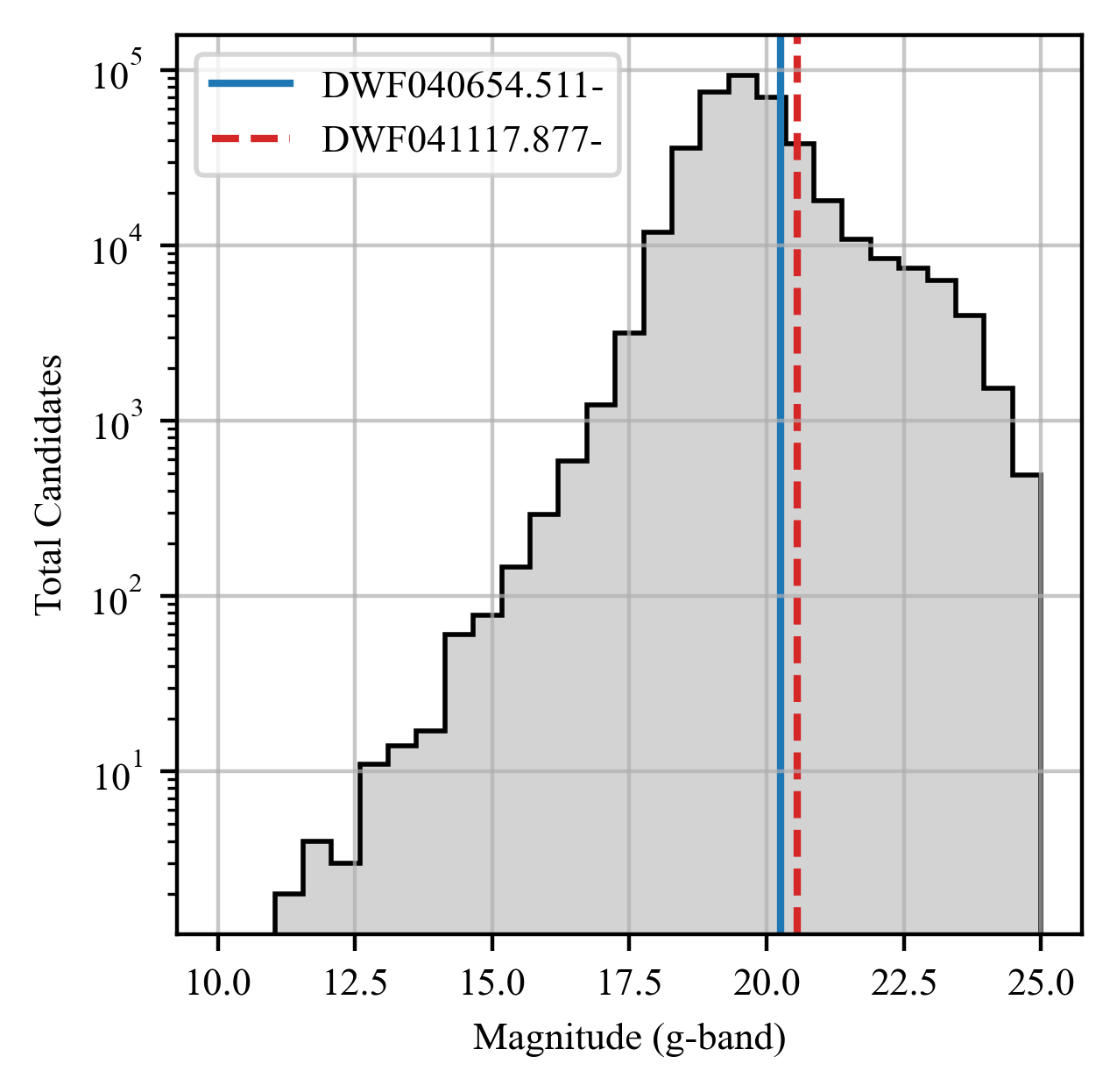}
\caption{Distribution of apparent \textit{g}-band magnitudes of all 385,775 single-detection candidates, highlighted with the distribution of the two promising candidates. The 5$\sigma$ limiting magnitude determined from all sources in the field is m($g$)$\sim$22.5-23}
\label{fig:mag_dist}
\end{figure}

Figure \ref{fig:mag_dist} shows the distribution of observed apparent \textit{g}-band magnitudes of all 385,775 sub-minute transient candidates. The distribution shows a distinct peak for sources with magnitudes between 19-20, before falling off towards the estimated 5$\sigma$ limiting magnitude of 22.5-23 determined for the images using all sources.

\begin{figure}
        \includegraphics[width=\columnwidth]{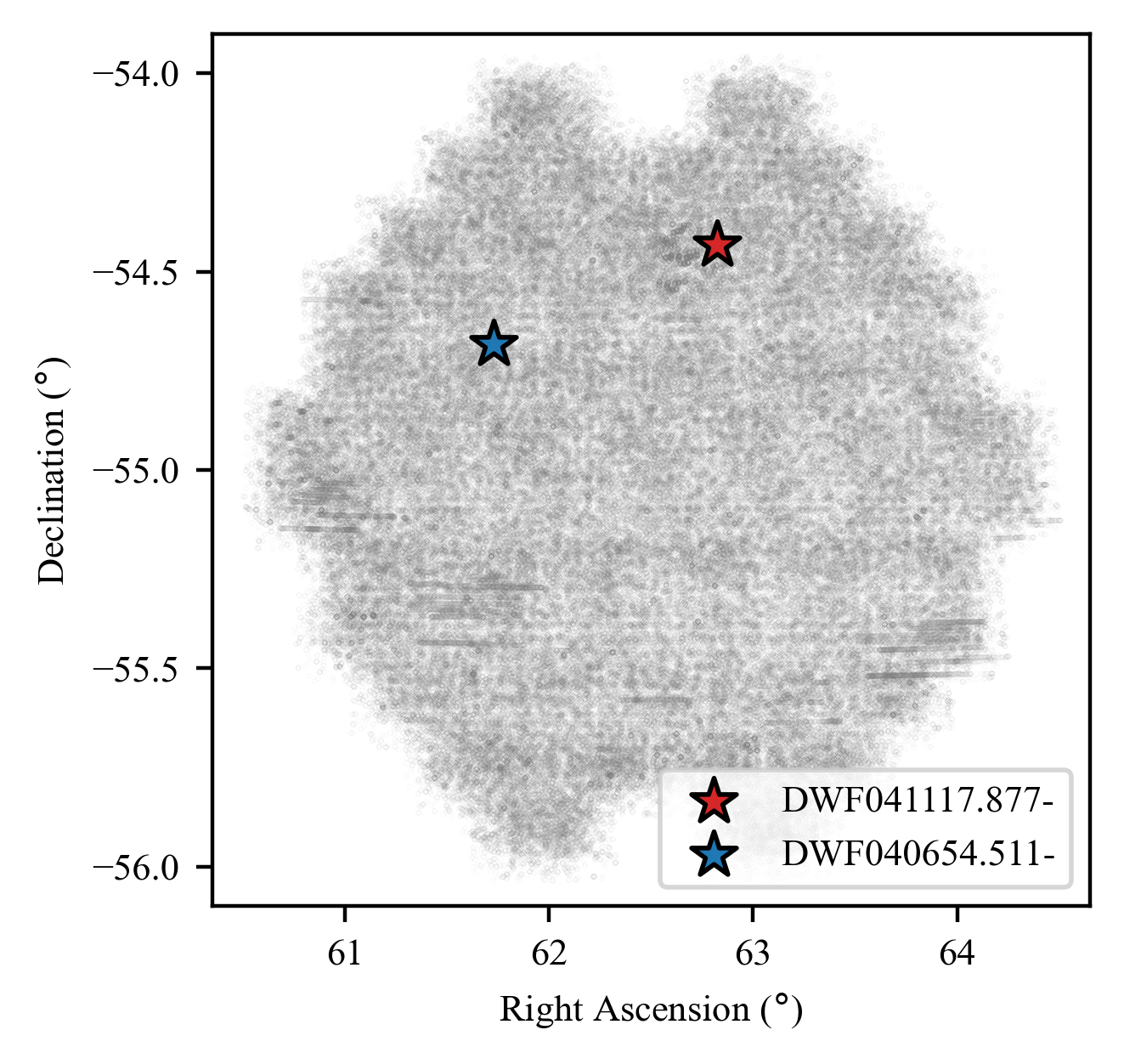}
        \caption{Sky positions of 4hr field candidates, with the two high-confidence candidates highlighted with red and blue stars. The two candidates have an angular separation of 0.68$^{\circ}$ and were detected roughly seven minutes apart. For both detections to be caused by a single moving object, said object would travel approximately 128 pixels during a single exposure. Both detections are PSF-like and do not streak, ruling out the moving-object hypothesis and suggesting astrophysical origins.}
        \label{fig:skyplot}
\end{figure}

Figure \ref{fig:skyplot} depicts the sky coordinates of all 385,775 candidates in the 4hr field, with the two promising sub-minute transient candidates highlighted.
Notably, there are no obvious signs of artefact behaviour near the candidates, such as an unusual number of candidates appearing in a single CCD or in lines along CCD edges or amplifiers.
When corroborated with Figure \ref{fig:skyplot}, Figure \ref{fig:xypos} demonstrates that the two promising candidates have sky and pixel locations consistent with random transient detection, not associated with CCD edge noise, hot pixels or other non-astrophysical phenomena.
We, therefore, conclude that these candidates are unlikely to be the result of electronic artefacts caused by CCD faults or imperfections.

\begin{table*}
    \begin{tabular}{ccccccccc}
        \hline
        Candidate Name & RA (hms) & Dec (dms) & MJD & \textsc{robot} & \textsc{fwhm} (px) & $\epsilon$ & \textsc{spread model} & \textit{g} mag\\ \hline
        DWF041117.877- & 4:11:19.368 & -54:26:07.534 & 57036.261479 & 0.930 & 6.09 & 0.054 & $4.89\times10^{-4}$ & 20.55 $\pm$ 0.04\\
        DWF040654.511- & 4:06:56.218 & -54:41:08.861 & 57036.265940 & 0.995 & 6.69 & 0.108 & $2.60\times10^{-5}$ & 20.25 $\pm$ 0.05 \\ \hline
        \end{tabular}
    \caption{Detection, location and source extraction information for the two high-confidence candidates.}
    \label{tab:candetails}
 \end{table*}
        
\begin{figure*}
    \centering
    \includegraphics[width=\linewidth]{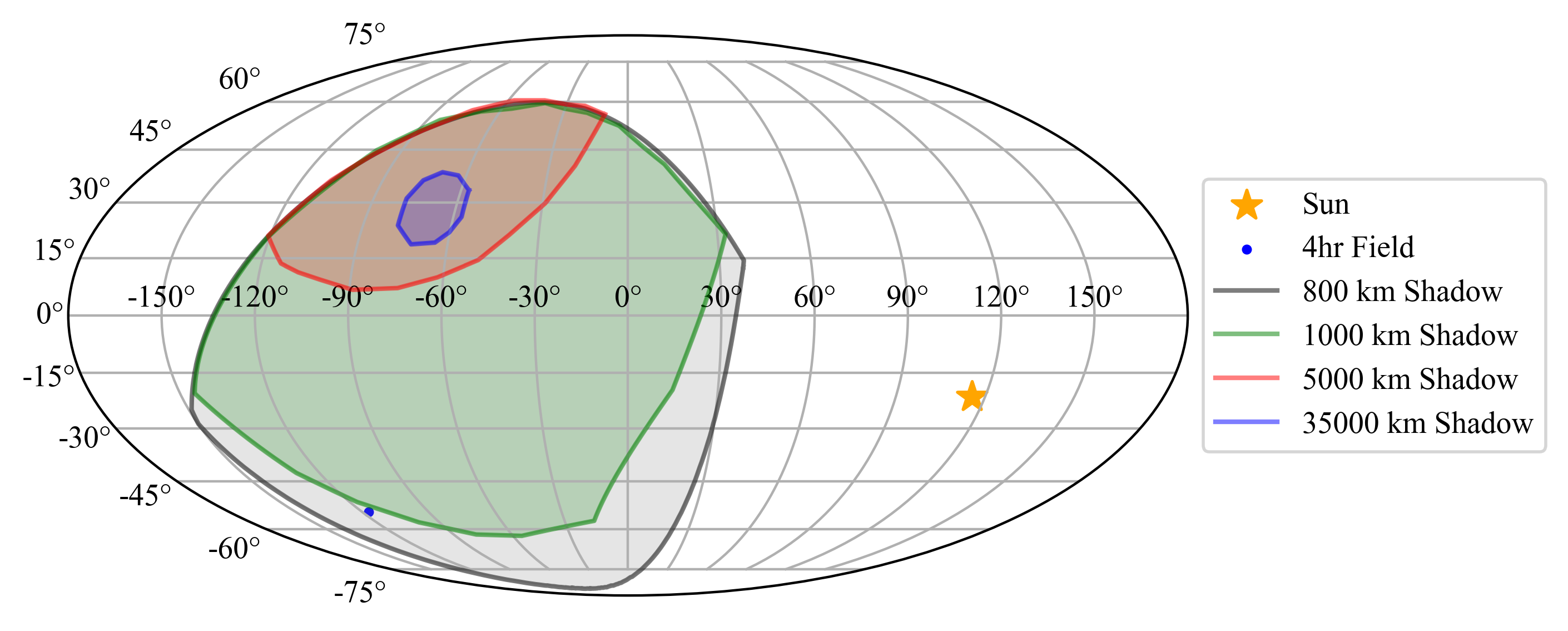}
    \caption{All sky view of the 4hr field where the two candidates were detected.
    The orange star is the position of the Sun at the time of the observations.
    The blue dot indicates the central location of the 4hr field.
    The shaded regions represent the Earth's umbra at various altitudes, where objects are shadowed and therefore do not reflect sunlight.
    During our observations, the 4hr field was within the 800km shadowed region.
    Therefore, we can confidently rule out any artificial satellites below an altitude of 800km as the progenitor of our transient candidates. Note that the shadows have been projected to the limit of the Earth's horizon from the CTIO location during the time of candidate observations.}
    \label{fig:Shadows}
\end{figure*}

\subsection{Multiwavelength Counterpart Search}
\label{subsec:multiwavelength}

As the 2015 DWF run assessed here was a pilot run for the program, only two other multiwavelength telescopes observed the target fields simultaneously with the DECam; namely, the MOST and the Murriyang (Parkes) radio telescopes.
The Neil Gehrels Swift Observatory was available for rapid-response target-of-opportunity observations for sources detected in the DECam, Murriyang or MOST observations in real-time.
We note here that MOST reported zero real-time transient alerts during these observations.

We are particularly interested in identifying any high-energy burst events in the minutes before and after the short-duration optical candidates identified in this work.
The archival data from the Neil Gehrels Swift Observatory was searched for any detections that coincided with the hard X-ray Burst Alert Telescope (BAT) instrument \citep{Barthelmy2005}.
Neither of our high-confidence candidates was in the BAT field of view immediately before or after the events and subsequently had zero detections. 

We used \texttt{Heimdall} single pulse software \citep{heimdall}\footnote{\href{https://sourceforge.net/projects/heimdall-astro/}{https://sourceforge.net/projects/heimdall-astro/}}, to search the Murriyang data for signals in 'gulps' typically 16.8 s in length with a dispersion measure tolerance of 1.05.
We then used \texttt{Your} \citep{YOUR} to produce dedispersed frequency-time and DM-time information for the \texttt{Heimdall} candidates.
We ran \texttt{FETCH} \citep{FETCH1, FETCH2}, a machine-learning algorithm, on these candidates to assign a probability of being an FRB to each, and visually inspected all candidates with a probability greater than 0.75.
In these searches, no high signal-to-noise (SNR $>$ 10) FRB candidates were found, and no evidence for a fainter FRB was found down to an SNR of $\sim$7.
We note that DWF040654.511- is located in the gap between beams 3, 4, and 10, and DWF041117.877- is just outside of beam 2.

\subsection{Rates}
\begin{figure}
    \centering
    \includegraphics[width=\columnwidth]{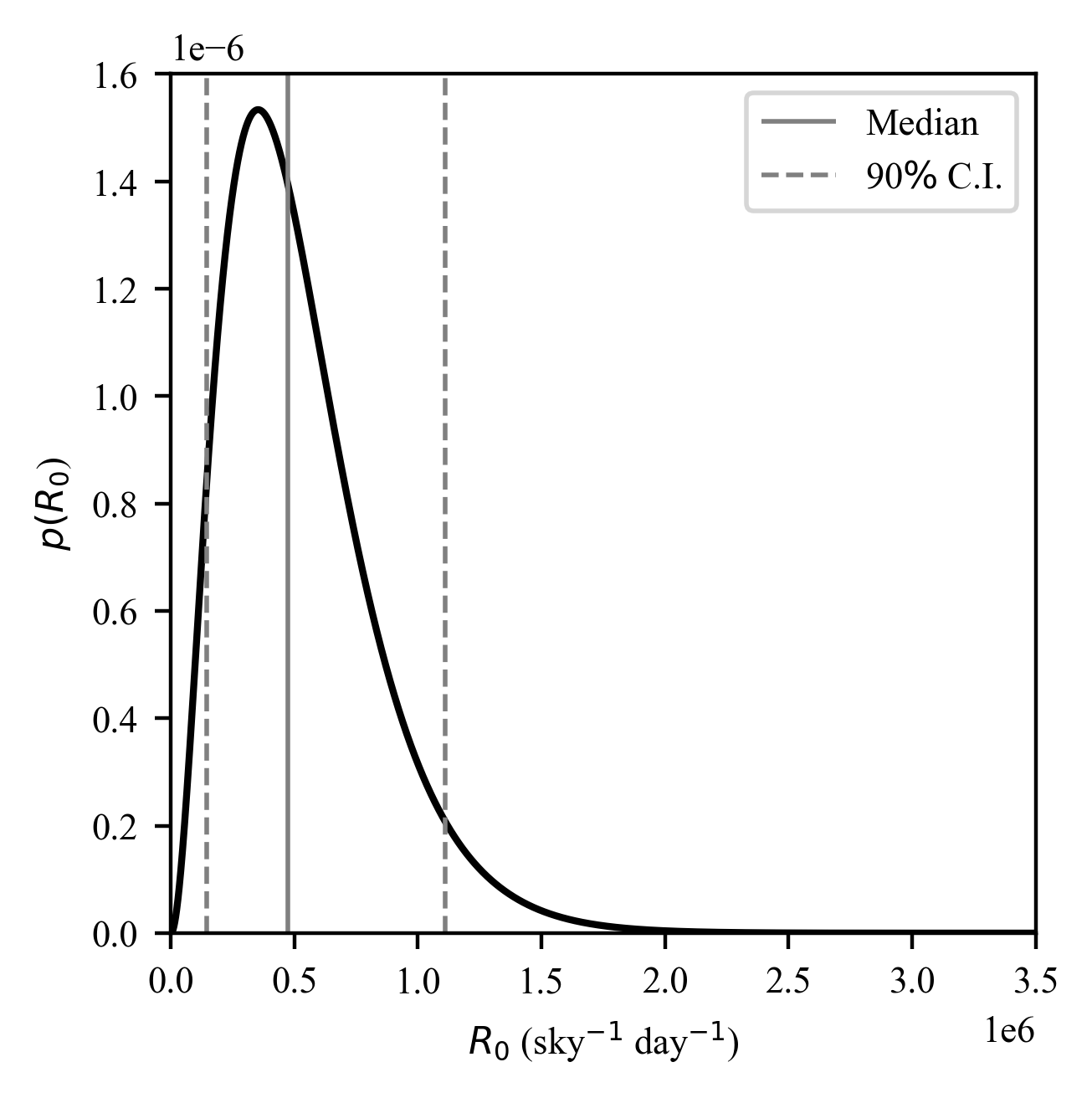}
    \caption{Probability density function of the rate of sub-minute optical transient events across the sky per day. From this distribution, we calculate a median rate of $4.72^{+6.39}_{-3.28}\times10^5$ events on the sky per day (90\% C.I.).}
    \label{fig:rates}
\end{figure}
We estimate the rates of sub-minute optical transient events with properties similar to those of the two identified candidates.
Naturally, the relatively short time on sky our data spans provides a poor lever arm for an accurate rate measurement.
We construct a probability density function for the rate of events $R_0$, using Poisson statistics,
\begin{equation}
    p(R_0)=\frac{(R_0t_{obs}\Omega)^ne^{-(R_0t_{obs}\Omega)}}{n!}
\end{equation}
where $t_{obs}$ is the total exposure time in seconds,
$\Omega$ is the field of view in deg$^{2}$
and $n$ is the number of detected events.
Figure \ref{fig:rates} shows the event rate probability density function scaled to units of events per day across the entire sky. 
From this distribution, we calculate the median and 90\% confidence intervals to be
\begin{equation}
    R_0=4.72^{+6.39}_{-3.28}\times10^5\text{ events on sky per day.}
\end{equation}

One interpretation, if the sources are astrophysical, of this high rate is that these sub-minute optical transient events may repeat.
While we leave later DWF DECam optical and the complete multiwavelength late-time counterpart search for future work, we did have accessible radio data to search for late-time radio counterparts to our two optical transient candidates.
We explore the possibility of connection to other radio events, including radio variable and slow transient sources, via additional available radio imaging.
The CSIRO's Australian SKA Pathfinder \citep[ASKAP\footnote{\href{https://www.atnf.csiro.au/projects/askap/index.html}{https://www.atnf.csiro.au/projects/askap/index.html}};][]{2021PASA...38....9H} is a 36-dish radio interferometer located at Inyarrimanha Ilgari Bundara, the CSIRO's Murchison Radio-astronomy Observatory.
ASKAP has a $\sim$30 square degree field of view, and uses that wide field of view to survey the sky at a range of frequencies and time scales.
All ASKAP data are publicly available in the CSIRO ASKAP Science Data Archive (CASDA)\footnote{\href{https://research.csiro.au/casda/}{https://research.csiro.au/casda/}}.

There are 39 and 43 ASKAP observations covering the positions of DWF040654.511-544056.411 and DWF041117.877-542554.144 respectively.
These observations occurred between 27 August 2019 and 27 June 2024 with integration times ranging from 12 min to 10 hours.
Neither source is detected in any ASKAP observation.
The image with the longest integration time for both sources is a 10-hour Evolutionary Map of the Universe \citep[EMU;][]{2011PASA...28..215N} observation at 943.5\,MHz taken on 2024 February 26 (observation ID: SB59481)
We used the \textsc{RadioFluxTools}\footnote{\href{https://gitlab.com/Sunmish/radiofluxtools}{https://gitlab.com/Sunmish/radiofluxtools}} package to perform forced flux density fitting at the positions of DWF040654.511-544056.411 and DWF041117.877-542554.144 in the EMU image.
The RMS value local to the sources was 0.03\,mJy, and we measured non-detection flux densities of $0.09\pm0.05$\,mJy and $0.05\pm0.04$\,mJy, respectively.
Both sources are covered by the Variables and Slow Transients with ASKAP \citep[VAST;][]{murphy2013} extragalactic survey, so will continue to be observed every couple of months over the next two years.

\subsection{Satellite debris glint rejection}
\label{sec:glints}

There are currently over 130 million pieces of space debris in orbit around the Earth larger than 1mm in size \citep{BONGERS2023}. Of these, only $\sim$27,000+ pieces of debris are actively tracked and catalogued by the North American Aerospace Defence Command\footnote{Correct as of 2025}. The majority of current space debris, estimated to be less than 1cm in size, poses a challenge to effective tracking \citep{Aglietti2020}. Currently, small optical and active radar tracking facilities are generally limited in their ability to detect debris of 1cm in low Earth orbit (LEO), leaving a significant gap in our knowledge of debris around the Earth \citep{Hamilton2017}. As the number of resident space objects grows, so does the possibility of satellite and debris glints being misidentified as astronomical transients. Therefore, we investigate the likelihood that these two candidates are not from astrophysical origins. 

To determine whether the two candidates could have resulted from satellite or space debris glints, we first identified the Earth's umbra eclipse factor, which determines if an object at a given altitude is in the Earth's umbral shadow at the given date and observation time \citep{Fixler1964, Ismail2015}. The eclipse factor can be calculated for varying altitudes above the Earth, with the angular radius of the Earth's (approximately circular) umbral shadow projected onto the sky decreasing with altitude. We calculated the umbra eclipse factor for LEO at altitudes of 800 km and 1,000 km, medium Earth orbit (MEO) at an altitude of 5,000 km, and geosynchronous orbit (GEO) at an altitude of 35,000 km, specifically at the CTIO location on Earth and during the time of our observations. This can be seen in Figure \ref{fig:Shadows}.

We determined that the 4hr field was not within the umbra eclipse region for objects above 1,000 km. This allows us to eliminate the possibility of sources arising from satellite or debris glints from objects with an orbit below 800 km. This does not rule out the possibility that a glint could have occurred at a higher altitude. However, the candidates do not display typical traits associated with objects in orbit. Both candidates' FWHM (6.09 and 6.69 pixels) are consistent with astrophysical point sources on the sky, as are the \texttt{SPREAD\_MODEL} values for objects of this magnitude from \textsc{source extractor}. Typical resident space objects in orbit around the Earth will exhibit streak-like shapes or tracklets \citep{Piattoni2014A, Virtanen2016, Karpov2023}. We visually investigated each CCD where the candidates were identified and found no evidence of streaks or tracklets. 

To investigate further whether the two candidates could have resulted from debris, we modelled the apparent visual magnitude of various sizes of resident space objects. This will validate if the magnitudes of the candidate sources could be consistent with detectable debris. For this work, we assume the objects are spherical and diffuse, and use the methods outlined by \citep{McCue1971} to calculate the apparent brightness with the following equation:

\begin{equation}
    m = -26.74 - 2.5 \log_{10} \left( \frac{A \gamma f(\phi)}{z^2} \right) + x \chi
\end{equation}
where 
\( A \) is the cross-sectional area, 
\( \gamma \) is the albedo or reflectivity, 
\( \phi \) is the solar phase angle, 
\( f(\phi) \) is a function that defines the fraction of reflected light based on the solar phase angle, 
\( z \) is the range of the satellite from the observer, 
\( x \) is the atmospheric absorption coefficient (0.12 mag/airmass in the V-band; see \citet{Patat2011, HainautWilliams2020}), and \( \chi \) is the airmass.
This is a conservative assumption, as a flat mirror-reflective surface would appear significantly brighter.

\begin{figure*}
    \centering
    \includegraphics[width=\linewidth]{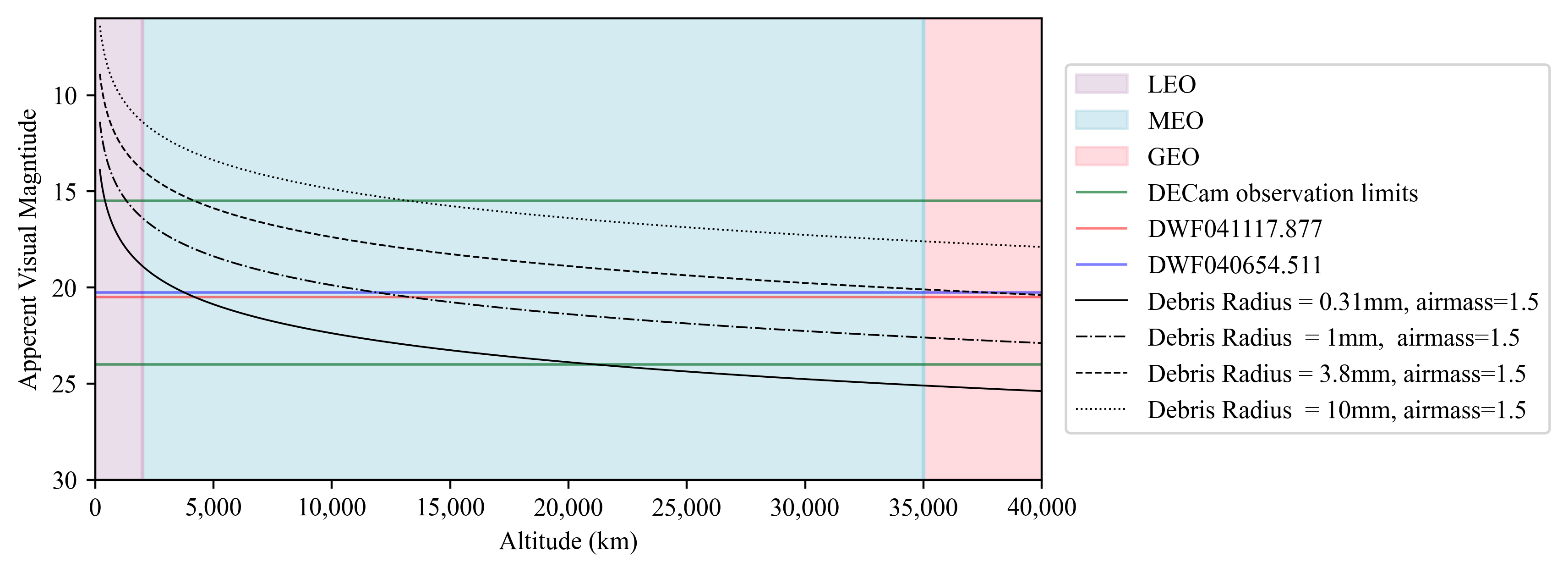}
    \caption{Modelling for satellite source rejection of the apparent visual magnitude of various sizes of debris, across altitude ranges. LEO, MEO and GEO orbits are represented by the purple, blue and red shading, respectively, and DECam saturation and limiting magnitudes are shown in grey horizontal lines. The two red horizontal lines are the magnitudes of the two candidate sources in this work.}
    \label{fig:satbrightness}
\end{figure*}

We assumed a 50\% fraction of reflected light to account for the low solar phase angle, given the time of night these observations occurred. The results are shown in Figure \ref{fig:satbrightness}. From this modelling, we can narrow down the most likely size of an object that could cause glints of a similar magnitude to the two candidates between a radius of $\sim$ 1mm and $\sim$ 4mm at altitudes above typical LEO. To further identify if this is a likely scenario, we estimated the rate of movement across the sky in 20 seconds of exposure time. This allowed us to determine how long the streak from reflected sunlight would appear at typical orbital speeds across the DECam CCDs. The streaks were calculated to span a minimum of $\gtrsim$900 pixels for objects in a MEO and LEO orbit. Again, as both candidates appear to be PSF-like, with a FWHM under $\sim$7 pixels, this is inconsistent with typical debris detection. We conclude that these two events are unlikely to be caused by space debris glints.

Finally, we use the image profiles of the two promising candidates as a strong constraint on the possibility of satellite glints and their possible distance.
Firstly, the durations of the sources are required to be longer than $\sim$1/50$^\text{th}$ of a second to enable passage of their light through the atmosphere to form a PSF \citep{Beavers1989}.
Secondly, their motion on the sky cannot be more than one pixel during the 20-second exposure to be consistent with the \textsc{spread\_model} profiles.
A satellite in a geostationary orbit moves at $\sim$15'' per second.
Thus, a glint from such a satellite would need to be $\lesssim$1/60$^\text{th}$ of a second to move one DECam pixel (0.263'' per pixel) or less.
Any satellite in a closer orbit would move too fast to form a complete PSF, and is ruled out.
Such a short glint from a geostationary satellite would unlikely form such a PSF as seen in Figure \ref{fig:2cands}.

\subsection{Possible progenitors}
\label{subsec:discussion}
We report two promising candidates that, when corroborated with the diagnostic plots in \S\ref{sec:results_and_analysis}, can potentially be sub-minute astrophysical optical transient events.
Figures \ref{fig:spreadmodel}, \ref{fig:xypos}, and \ref{fig:skyplot} all provide evidence to suggest these two candidates are not the result of CCD-related faults.
Section \ref{sec:glints} indicates these two candidates are unlikely to be glints from resident space objects in orbit around the Earth.
We propose that these two candidates are of astrophysical origin.
Further supporting this is candidate DWF040654.511-544056.411, which may have an associated host galaxy in deep imaging.
DWF040654.511-544056.411 will be analysed in further detail in a future publication (Goode et al., in preparation).
In addition, candidate DWF041117.877-542554.144 does not have a visible host galaxy in available deep imaging and could be related to a Galactic event.

We conducted a query through the Transient Name Server to search for any transient records associated with our two candidates within a 50" search radius, which returned no results.

Here, we discuss astrophysical or artificial phenomena that could explain the candidates. The first possibility is that these events result from beta decay within the high-potassium glass of the dewar within the DECam instrument. This type of decay is usually associated with worm-like tracks, and not consistent with the candidates identified in \cite{Smith2002}. Another possibility is that these two events were caused by cosmic-ray muons, coined `spots' in \cite{Smith2002}; however, spots typically do not have a PSF-like appearance and were one of the artefacts trained explicitly within the machine-learning pipeline and lends confidence that the sources are unlikely due to detector or decay artefacts. 

A variety of extragalactic transient events are thought to be associated with short-duration (minutes to days) optical bursts.
There has been evidence for optical emission resulting from type II-P supernovae breakouts reported by \cite{Garnavich2016}.
Luminous optical flashes accompanying Gamma Ray ($\gamma$-ray) Bursts (GRBs) have been observed with fade rates in excess of 1 magnitude per minute \citep[e.g.][]{Oganesyan2023}.
There have also been multiple studies searching specifically for optical counterparts to FRBs since their discovery.
Optical bursts from repeating FRBs have also been searched for and upper limits on their brightness have been constrained previously \citep{Hardy017,Tingay2022,Kilpatrick2024}.
No searches for counterparts to single-burst FRBs have been undertaken.
These searches can only be accomplished with simultaneous multiwavelength observations, such as the DWF programme.

These transients are accompanied, or theorised to be accompanied, by other transient behaviours or multiwavelength counterparts.
Namely, type Ia or II shock breakouts quickly evolve into their respective supernovae, GRB optical flashes are expected to be accompanied by prompt, $\gamma$-ray and x-ray emission, and FRB counterparts, which are associated with FRBs and may occur before, simultaneously or after the FRB.
Recently discovered, AT2022tsd is a Luminous Blue Optical Transient (LFBOT) and an AT2018cow-like transient that exhibited minutes-duration optical flares of supernova-like magnitudes \citep{Ho2023}.
LFBOTs with these characteristics serve as a possible explanation for the progenitor of extragalactic sub-minute transients, with a magnetar or accreting black hole as an embedded energy source.
As telescope surveys can only detect a fraction of the transients in the sky, it becomes challenging to identify the exact nature of our sub-minute optical transient candidates.
A further study into the possible progenitors and mechanisms for our candidates will be explored in future work.

Galactic transients have also been observed to occur on short timescales.
\citet{Webb2021}, working with DWF DECam imaging data, found that stellar flares can occur on timescales as short as 3 minutes, limited only by the selection criteria applied to their search.
This suggests that stellar flares may exist on even shorter timescales, including in the sub-minute regime.
This work presents one sub-minute optical transient candidate with no apparent host, DWF041117.877-542554.144, and the other may not be associated with its nearby galaxy.
We conducted a query through \textit{Vizier} to search for additional information on the source but found no records from any surveys within 8.5" of the source.
As such, we believe the source is more likely to have Galactic origins rather than an unseen host galaxy.
With deep, longer wavelength imaging, it may be possible to conclude that the source is a stellar flare and provide some detail on its progenitor.
Given the magnitude of the source and the lack of an apparent host, it remains the prevailing, hypothetical explanation for this sub-minute optical transient candidate.

\subsection{False-positive explanations}
We have provided reasonable evidence that precludes these two sub-minute transient candidates from being caused by CCD-related electronic artefacts, and we have suggested some possible explanations for their origins.
However, we wish to elaborate further on possible false-positive cases.


As mentioned at the beginning of this section, manual inspection of the data revealed several misclassified candidates, most of which are suspected to be anomalous cosmic rays.
These anomalous cosmic rays typically presented as sharp, minuscule dots (some as small as 2$\times$2 pixels) with no trailing tails or streaking patterns.
Although these artefacts are readily identifiable by eye, the \textsc{robot} CNN misclassifies these types of cosmic rays because no such objects were available in its training dataset.
These objects have appeared to be the only meaningful data processing discrepancy between the \textsc{robot} training data (\textit{Mary} pipeline; \citet{Andreoni2017}) and the candidates in this work (NOAO community pipeline; \citet{Valdes2014}).
These misclassifications by \textsc{robot} indicate that we cannot be totally reliant on \textsc{robot}'s assessment, and we should be cautious of its biases; however, measurements from \textsc{source extractor} coupled with human inspection rule out these candidates from the final set of promising candidates.


We explore the possibility that these two candidates may be flashes of reflected sunlight, briefly shining from a reflective surface of a spinning artificial satellite or debris.
In the exposures immediately before and after these candidates were detected, we found no similar sources in the nearby vicinity on-sky, indicating that they do not blink rapidly enough to be tracked in our data.
In addition, we find that the detection of these candidates occur roughly at 6:16 AM (UTC) and 6:23 AM (UTC) (for DWF041117.877- and DWF040654.511- respectively) which, when converted into the local timezone at CTIO (CLST at the observing time), indicates that they occurred at approximately 3:16 AM and 3:23 AM, respectively.
Modelling, detailed in Section \ref{sec:glints}, determined that the 4hr field was within the Earth's shadow for altitudes below 800 km, ruling out these objects as progenitors. However, the field was not shadowed for objects above 1,000 km, meaning they would have been sunlit and remain a possibility.
However, their detection properties are inconsistent with previously detected debris objects (i.e. no evidence of streaks or tracklets).

While we have ruled out certain phenomena as the source of the two promising candidates, there is insufficient information to decisively conclude upon the true nature and origins of these sub-minute events.
However, our work suggests the candidates may be astrophysical, and several characteristic observations can be made.
Candidate DWF040654.511-544056.411 is not strongly associated with an extragalactic host and is believed to be a Galactic transient, namely sub-minute stellar flares, where the source star is fainter than our detection threshold.
Meanwhile, DWF041117.877-542554.144 does appear to have a potential host galaxy, which will be explored in depth in future work. 

\section{Summary}
\label{sec:conclusions}
In this work, we showcase a novel, machine-learning-enabled method to search for optical transients on sub-minute timescales using the DWF programme.
The sub-minute optical transient discovery pipeline aims to maximise the scientific impact of data that is traditionally large and heavily contaminated with artefacts; lightcurves that contain a single detection.
We processed 671,763 light curves from early DWF data, which covered two fields (4hr \& CDF-S) over 4 nights, for a total exposure time of 127 minutes (73 and 54 minutes, respectively).
During processing, the pipeline identified 385,775 light curves that contained a single detection, representing the majority of all available data (57.4\%).
Each of these 385,775 sources then had templates and subtractions created for visual inspection, \textsc{source extractor} parameter measurement and assessment by the \textsc{robot} Deep Convolutional Neural Network real/bogus classifier.
The \textsc{robot} CNN was used successfully to filter $>$98\% of single-detection light curves to a set of 5,477 sub-minute optical transient candidates.
Manual inspection of the 5,477 candidates revealed the vast majority to be contaminants, caused by using a real/bogus decision boundary of 0.06 (a high false negative rate), or misclassifications caused by a class of artefact unseen in the \textsc{robot} algorithm's training data. 

This work presents two candidates with PSF-like appearance and are plausible sub-minute optical transients. We provide evidence to suggest these candidates are not caused by cosmic rays or CCD-related electronic faults (Fig. \ref{fig:spreadmodel}, \ref{fig:xypos} \& \ref{fig:skyplot}). One candidate appears to have a galaxy-like object projected near the line of sight of the transient, potentially the candidate host galaxy. In addition, we report on a single candidate that has a strong PSF-like appearance that does not appear to have a host, nor any additional information from other wavelength regimes. We suggest the latter may have Galactic origins. Finally, we estimate the rates of sub-minute optical transient events to be $4.72^{+6.39}_{-3.28}\times10^5$ events on the sky per day.

In the lens of the upcoming Vera C. Rubin Telescope's Legacy Survey of Space and Time (LSST), we normalise these rates to an expected quantity observed per night by the survey.
By establishing an expected frequency, these rates assist brokers (e.g. ALeRCE \citep{Forster2021}, AMPEL \citep{Nordin2019}, ANTARES \citep{Matheson2021}, Babamul\footnote{\href{https://github.com/babamul/babamul}{https://github.com/babamul/babamul}}, Fink \citep{Moller2021}, Lasair \citep{Williams2024}) in optimising their real-time filtering, classification, and follow-up strategies for the millions of LSST alerts anticipated nightly.
We assume a field of view of 9.6 deg$^2$, and an exposure time of 30 seconds.
We also assume the survey will target 200 fields per night, with one exposure in the $g$-band each.
We note that this rate does not account for differences in depth or selection criteria between DECam and Rubin.
We normalise the rates by multiplying through a scaling factor, $F$, which accounts for the relative area and exposure time of one night of observations.
We define the scaling factor as
\begin{align*}
    F &= \text{relative area} \times \text{relative exposure time} \\
    F &= \frac{(200\times9.6)}{41,253} \times \frac{30}{86,400} \\
    F &= 1.6\times10^{-5}
\end{align*}
We calculate the scaled rate of sub-minute optical transients for LSST to be
\begin{equation}
    R_{LSST} = 7.6^{+10.3}_{-5.3} \text{ events per night}
\end{equation}

\section*{Acknowledgements}

Part of this research was funded by the Australian Research Council Centre of Excellence for Gravitational Wave Discovery (OzGrav), CE170100004 \& CE230100016. JC acknowledges funding by the Australian Research Council Discovery Project, DP200102102. AM is supported by DE230100055.

This project used data obtained with the Dark Energy Camera (DECam), which was constructed by the Dark Energy Survey (DES) collaboration. Funding for the DES Projects has been provided by the U.S. Department of Energy, the U.S. National Science Foundation, the Ministry of Science and Education of Spain, the Science and Technology Facilities Council of the United Kingdom, the Higher Education Funding Council for England, the National Center for Supercomputing Applications at the University of Illinois at Urbana-Champaign, the Kavli Institute of Cosmological Physics at the University of Chicago, the Center for Cosmology and Astro-Particle Physics at the Ohio State University, the Mitchell Institute for Fundamental Physics and Astronomy at Texas A\&M University, Financiadora de Estudos e Projetos, Fundação Carlos Chagas Filho de Amparo à Pesquisa do Estado do Rio de Janeiro, Conselho Nacional de Desenvolvimento Científico e Tecnológico and the Ministério da Ciência, Tecnologia e Inovacão, the Deutsche Forschungsgemeinschaft, and the Collaborating Institutions in the Dark Energy Survey. The Collaborating Institutions are Argonne National Laboratory, the University of California at Santa Cruz, the University of Cambridge, Centro de Investigaciones Enérgeticas, Medioambientales y Tecnológicas-Madrid, the University of Chicago, University College London, the DES-Brazil Consortium, the University of Edinburgh, the Eidgenössische Technische Hochschule (ETH) Zürich, Fermi National Accelerator Laboratory, the University of Illinois at Urbana-Champaign, the Institut de Ciències de l’Espai (IEEC/CSIC), the Institut de Física d’Altes Energies, Lawrence Berkeley National Laboratory, the Ludwig-Maximilians Universität München and the associated Excellence Cluster Universe, the University of Michigan, the National Optical Astronomy Observatory, the University of Nottingham, the Ohio State University, the OzDES Membership Consortium the University of Pennsylvania, the University of Portsmouth, SLAC National Accelerator Laboratory, Stanford University, the University of Sussex, and Texas A\&M University.

This study is based on observations at Cerro Tololo Inter-American Observatory, National Optical Astronomy Observatory which is operated by the Association of Universities for Research in Astronomy (AURA) under a cooperative agreement with the National Science Foundation.

Murriyang, CSIRO’s Parkes radio telescope, is part of the Australia Telescope National Facility\footnote{\href{https://ror.org/05qajvd42}{https://ror.org/05qajvd42}} which is funded by the Australian Government for operation as a National Facility managed by CSIRO. We acknowledge the Wiradjuri people as the Traditional Owners of the Observatory site.

This scientific work uses data obtained from Inyarrimanha Ilgari Bundara, the CSIRO Murchison Radio-astronomy Observatory. We acknowledge the Wajarri Yamaji People as the Traditional Owners and native title holders of the Observatory site. CSIRO’s ASKAP radio telescope is part of the Australia Telescope National Facility$^6$. Operation of ASKAP is funded by the Australian Government with support from the National Collaborative Research Infrastructure Strategy. ASKAP uses the resources of the Pawsey Supercomputing Research Centre. Establishment of ASKAP, Inyarrimanha Ilgari Bundara, the CSIRO Murchison Radio-astronomy Observatory and the Pawsey Supercomputing Research Centre are initiatives of the Australian Government, with support from the Government of Western Australia and the Science and Industry Endowment Fund.

\section*{Data Availability}

The data used in this work are available at a reasonable request to the corresponding author.
All ASKAP data are publicly available in the CSIRO ASKAP Science Data Archive (CASDA)\footnote{\href{https://research.csiro.au/casda/}{https://research.csiro.au/casda/}}.
The source code of the sub-minute optical transient event discovery pipeline is available on GitHub\footnote{\href{https://github.com/simongoode/SMOTEs}{https://github.com/simongoode/SMOTEs}}.



\bibliographystyle{mnras}
\bibliography{full.bib} 




\appendix

\section{Unsupervised learning search for Sub-Minute Optical Transients with \texttt{HDBSCAN}}
\label{app:SE_robot}

This work explored the use of unsupervised methods to isolate and identify likely real astrophysical sub-minute optical burst candidates. To do this, we use feature extraction across both the light curves and the CCD images. Specifically, we utilise \textsc{source extractor} to calculate and measure the statistics of each source. In particular, the source extractor features are \textsc{class\_star, ellipticity, spread\_model} and \textsc{fwhm}, as measured from the science image. These values are collected for each candidate. Refer to \citet{Bertin1996} for details and definitions of \textsc{source extractor} parameters.

In addition to the features from \textsc{source extractor}, we use the real/bogus classifier within the \textsc{robot} pipeline \citep{Goode2022}. The real/bogus classifier is a deep convolutional neural network \citep[CNN;][]{LeCun1998,Lecun2015} trained on 31$\times$31 pixel resolution cutouts of the template, science and subtraction images. The algorithm was trained on 10,000 samples from a variety of DECam/DWF data and was designed to be particularly robust at classifying bogus samples (e.g. artefacts), so as to be conservative when classifying real samples. For more details on the architecture, data, construction and deployment of the CNN algorithm, refer to \citet{Goode2022}. The feature we used from this process is the real/bogus score for each candidate. The seven features extracted and used throughout this work can be seen in Table \ref{tab:features}.

\begin{table*}
\begin{tabular}{ll}
\hline
Feature         & Description                                                                    \\ \hline
\textit{g}-band magnitude       & The magnitude of the source identified in the light curve point                    \\
\textit{g}-band magnitude Error & The magnitude error associated with the source identified in the light curve point \\
Class Star      & Source extractor class star value for source                                   \\
Ellipticity     & Source extractor ellipticity measure of source                                 \\
Spread Model    & Source extractor spread model measure of source                                \\
FWHM            & Source extractor full width half maximum measurement of source                 \\
\textsc{Robot} Score     & \textsc{Robot} real/bogus score from deep CNN.                                          \\ \hline
\end{tabular}
\caption{Features extracted using original light curves, source extractor and the \textsc{robot} pipeline.}
\label{tab:features}
\end{table*}

\subsection{Sub-minute Transient Candidate Selection \& Artefact Filtering}
\label{app:filtering}
\begin{figure*}
\centering
\includegraphics[width=\textwidth]{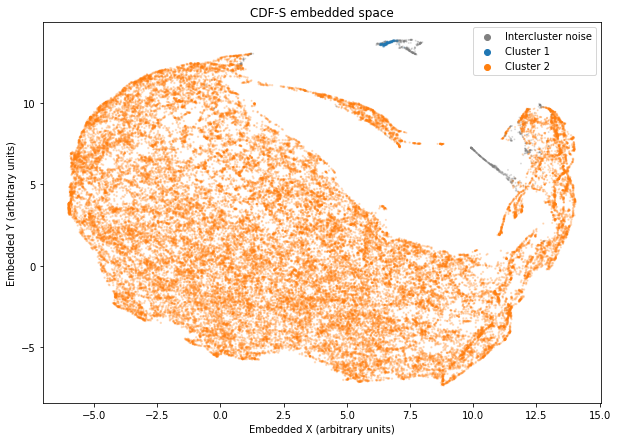}
\caption{UMAP embedded space of 46,646 candidates from the CDF-S field, highlighted with clustering from the HDBSCAN algorithm. Cluster 1 has been found to represent bright moving objects that streak across the field of view, and Cluster 2 represents a broad range of non-astrophysical artefacts and cosmic rays. The embedded space is a dimensionality reduction of the original, 7-dimensional feature space with arbitrary units in x and y, used only for visualisation.}
\label{fig:smotes_clustering}
\end{figure*}
\begin{figure}
\centering
\includegraphics[width=\columnwidth]{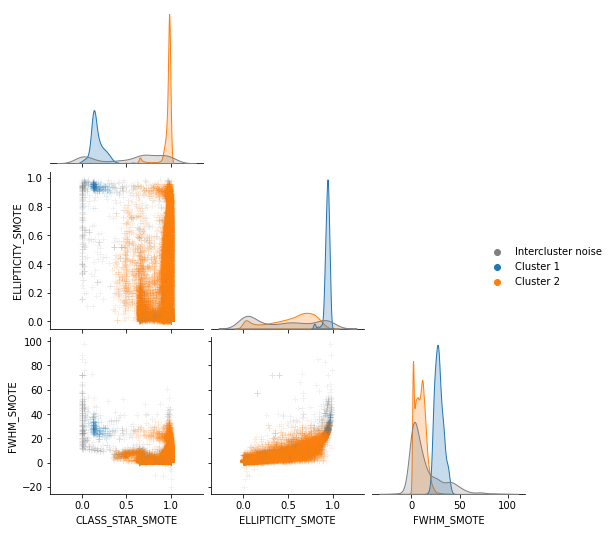}
\caption{Corner plot of high-impact features used for HDBSCAN clustering (\textsc{robot} score separated for readability; see Figure \ref{fig:cluster_robot}).}
\label{fig:cluster_features}
\end{figure}
\begin{figure}
\centering
\includegraphics[width=\columnwidth]{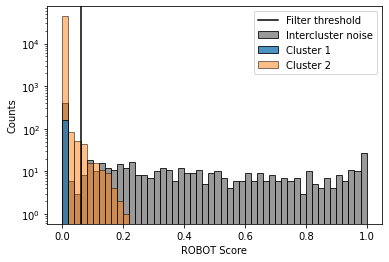}
\caption{Distribution of \textsc{robot} scores amongst HDBSCAN clusters. Clusters 1 and 2 contain exclusively low \textsc{robot} scores, indicating that intercluster noise will likely contain the highest quality candidates. A filter threshold of 0.06 was chosen to strike a balance between including as many intercluster noise samples as possible without including too many contaminants from Cluster 2.}
\label{fig:cluster_robot}
\end{figure}
\begin{figure*}
\centering
\includegraphics[width=\textwidth]{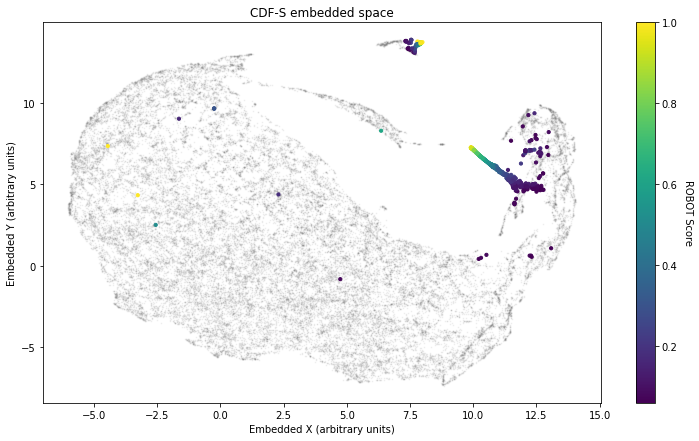}
\caption{UMAP embedded space of 46,646 candidates from the CDF-S field, highlighted with \textsc{robot} scores greater than 0.06. We find that the high-scoring candidates (i.e. those with scores close to 1) exist predominantly in the intercluster noise of Figure \ref{fig:smotes_clustering} and are the most likely candidates to be optical transients.}
\label{fig:smotes_robot_filtering}
\end{figure*}

In this work, we use unsupervised clustering methods to isolate regions of feature parameter space with the highest likelihood of containing sub-minute astrophysical candidates. We first employ the use of Hierarchical Density-Based Spatial Clustering of Applications with Noise \citep[HDBSCAN;][]{mcinnes2017hdbscan}. Clustering algorithms such as HDBSCAN attempt to identify similarities between data points with high dimensionality by mapping distances between points (distance metrics may vary, but this work utilises Euclidean distances). Data points that are small distances from each other tend to be similar to each other and thus form a cluster. With HDBSCAN, deciding how close points need to be to one another to form a cluster is decided dynamically. HDBSCAN will first attempt to find and cluster regions of high density before searching for lower-density clusters. Other parameters that affect the clustering, such as the minimum number of points required to form a cluster, are manually dictated by HDBSCAN hyperparameters.

For this approach, we utilise the features described in Appendix \ref{app:SE_robot}. Here, we strive to find PSF-like point source transients, the signal we expect from a genuine astrophysical transient. These combined features assess different aspects of the shape of a source and can be used to judge their PSF-like qualities. We omit subtraction image-based features, as not all samples produced a residual that could be measured by \textsc{source extractor}. Such samples produce \texttt{NaN} values and cannot be used in clustering. 

Visualising the 7-dimensional feature space to identify clusters is difficult. However, there are dimensionality reduction techniques available that assist with this problem. In this work, we utilise Uniform Manifold Approximation and Projection for Dimensionality Reduction \citep[UMAP;][]{McInnes2018} to reduce the actual 7-dimensional features into 2-dimensional embedded features. We note that the 2-dimensional embedded space uses arbitrary units and is only used as a visualisation aid.

We first experimented with these techniques on the CDF-S field candidates. 
Figure \ref{fig:smotes_clustering} shows the CDF-S field data as projected in its embedded space, showcasing the complex structure and relationships between samples. The coloured regions of this space indicate the clusters identified by HDBSCAN, including intercluster noise. The features that influence the clusters the most are \textsc{class\_star, ellipticity} and \textsc{fwhm}. Figure \ref{fig:cluster_features} shows how the features interact and how the clusters segregate based on impactful features. The HDBSCAN algorithm produces 2 distinct clusters that cover a wide expanse of the feature space.

Cluster 1 represents a region of feature space that is distinct from other data, and shows samples of high ellipticity, \textsc{fwhm} and low probability of being a star. From Figure \ref{fig:cluster_robot} we also identify this cluster to have, exclusively, very low \textsc{robot} scores. While these \textsc{source extractor} parameters suggest that these sources could be galaxies, the \textsc{robot} scores indicate that this cluster should consist of artefacts. Inspection of samples from this cluster revealed that the cluster represents streaks from bright and fast-moving objects across the field of view. In the CDF-S data, this cluster contains 163 samples and represents approximately 0.36\% of all candidates.

Cluster 2 contains the vast majority, encompassing 44,330 samples, or approximately 97.77\% of all candidates. This cluster encompasses samples with high \textsc{class\_star} values, low-to-moderate \textsc{fwhm} values, and a broad range of \textsc{ellipticities}. From Figure \ref{fig:cluster_robot} we once again identify a large proportion of this cluster to have very low \textsc{robot} scores, with only a handful of samples reaching as high as 0.2. From these values, we expect this cluster to represent a wide variety of artefacts, including electronic and photometric artefacts and cosmic rays. Inspection of samples from this cluster revealed a number of such artefacts, including cosmic rays of various intensities, shapes, and sizes, sky noise fluctuations, amplifier artefacts and crosstalk.

Finally, we have the remaining 849 samples ($\sim$1.87\% of all candidates) that did not fit into either cluster and which form the intercluster noise. Figure \ref{fig:smotes_clustering} shows small regions of intercluster noise dotted nearby the edges of Clusters 1 \& 2, as well as more prominent and populated regions that span further from the main clusters. We expect the intercluster noise to contain the most anomalous, and by extension, rarest candidates. Figure \ref{fig:smotes_robot_filtering} highlights the regions of the embedded space that contain high \textsc{robot} scores, colouring samples with values ranging from 0.06 to 1.0. This figure shows that the high-scoring \textsc{robot} candidates fall into the prominent regions of the intercluster noise, affirming that the most promising sub-minute optical candidates are likely to exist in intercluster noise. With a decision boundary at 0.06, Figure \ref{fig:robot_boundary} shows the real/bogus algorithm to have a false negative rate (FNR) of 0.6\%. Although such a low decision boundary will introduce contaminants, it is preferable to minimise the FNR as much as is affordable to be as conservative as possible.

Ultimately, we must devise an efficient method to filter out poor candidates to make manual inspection feasible. HDBSCAN clustering reveals that over 98\% of all candidates are artefacts and can effectively filter out the vast majority of uninteresting candidates by isolating the anomalous, unclustered samples. Using a real/bogus algorithm can further enhance filtration efficiency by removing the contaminants mentioned above, even at a conservatively low decision boundary. 

In the CDF-S field alone, the real/bogus algorithm finds 605 candidates above a score of 0.06, compared to 849 from intercluster noise. This marks a $\sim$40\% reduction in candidate numbers with the added confidence that most, if not all, high-quality candidates are included based on a 0.6\% FNR. 

We ultimately chose to apply this \textsc{robot} score threshold to the entire dataset of 385,775 candidates, which successfully filtered out the vast majority of candidates, leaving 5,477 with a higher likelihood of containing quality candidates. 
Clustering techniques are an interesting avenue to explore in future work, where real/bogus scores for transient candidates are either unavailable or unreliable. 


\bsp	
\label{lastpage}
\end{document}